\documentclass[prl,aps,twocolumn]{revtex4-1}
\usepackage{amsmath}
\usepackage{amssymb}
\usepackage{graphicx}
\usepackage{bbold}

\newcommand{\beqa}{\begin{eqnarray}}
\newcommand{\eeqa}{\end{eqnarray}}

\begin{document}
\title{$\mathcal{PT}$ restoration via increased loss-gain in $\mathcal{PT}$-symmetric Aubry-Andre model} 
\author{Charles Liang}
\author{Derek D. Scott}
\author{Yogesh N. Joglekar}
\affiliation{Department of Physics, 
Indiana University Purdue University Indianapolis (IUPUI), 
Indianapolis, Indiana 46202, USA}
\date{\today}
\begin{abstract}
In systems with ``balanced loss and gain'', the $\mathcal{PT}$-symmetry is broken by increasing the non-hermiticity or the loss-gain strength. We show that finite lattices with oscillatory, $\mathcal{PT}$-symmetric potentials exhibit a new class of $\mathcal{PT}$-symmetry breaking and restoration. We obtain the $\mathcal{PT}$ phase diagram as a function of potential periodicity, which also controls the location complex eigenvalues in the lattice spectrum.  We show that the sum of $\mathcal{PT}$-potentials with nearby periodicities leads to $\mathcal{PT}$-symmetry restoration, where the system goes from a $\mathcal{PT}$-broken state to a $\mathcal{PT}$-symmetric state as the average loss-gain strength is increased. We discuss the implications of this novel transition for the propagation of a light in an array of coupled waveguides.
\end{abstract}
\maketitle

\noindent{\it Introduction.} Open systems with balanced loss and gain have gained tremendous interest in the past three years since their experimental realizations in optical~\cite{expt1,expt2,expt3,expt4}, electrical~\cite{rc}, and mechanical~\cite{pendulum} systems. Such systems are described by non-hermitian Hamiltonians that are invariant under combined parity and time-reversal ($\mathcal{PT}$) operations~\cite{bender3}. Apart from their mathematical appeal, such non-hermitian Hamiltonians show non-intuitive properties such as unidirectional invisibility~\cite{expt4,uni1,uni2} and are thus of potential technological importance.  

Historically, $\mathcal{PT}$ Hamiltonians on an infinite line were the first to be investigated~\cite{bender1,bender2}. The range of parameters where the spectrum of  the Hamiltonian is purely real, $\epsilon_\lambda=\epsilon^*_\lambda$, and the eigenfunctions are simultaneous eigenfunctions of the $\mathcal{PT}$ operator, $\psi_\lambda(x)=\psi^*_\lambda(-x)$, is called the $\mathcal{PT}$-symmetric phase. The emergence of complex conjugate eigenvalues when the parameters are not in this region is called $\mathcal{PT}$-symmetry breaking. A positive threshold for $\mathcal{PT}$-symmetry breaking implies that the system transitions~\cite{expt2,expt3,expt4,rc,pendulum} from a quasi-equilibrium state at a small but nonzero non-hermiticity, to loss of reciprocity as the strength of the balanced loss-gain term crosses the threshold. Although $\mathcal{PT}$-symmetric Hamiltonian studies started with continuum Hamiltonians, all of their realizations are in finite lattices where the continuum, effective-mass approximation may not apply. This observation has led to tremendous interest in $\mathcal{PT}$-symmetric lattice models~\cite{znojil1,znojil2,bendix,song,avadh,longhi1,longhi2} that can be realized in coupled waveguide arrays~\cite{review1,review2}. 

A universal feature of all such systems is that $\mathcal{PT}$-symmetry is broken by increasing the balanced loss-gain strength and restored by reducing it. Here, we present a tight-binding model that can exhibit exactly opposite behavior, via a family of $\mathcal{PT}$-symmetric, periodic potentials. 

A remarkable property of lattice models, absent in the continuum limit, is the effects of a periodic potential. The spectrum of a charged particle in constant magnetic field in two dimensions consists of Landau levels~\cite{ll,graphene1,graphene2}; a similar particle on a two-dimensional lattice displays a fractal, Hofstadter butterfly spectrum~\cite{harper,butterfly,ca,graph}. In one dimensional lattices, a fractal spectrum emerges in the presence of a hermitian, periodic potential, and this model, known as the Aubry-Andre model~\cite{aa}, shows localization transition in a clean system when the strength of the incommensurate potential exceeds the nearest-neighbor hopping~\cite{lahini}. Here, we consider a $\mathcal{PT}$-symmetric Aubry-Andre model on an $N$-site lattice with hopping $J$ and complex potential $V_{\beta}(n)=V_0\cos\left[2\pi\beta(n-n_c)\right]+i\gamma\sin\left[2\pi\beta(n-n_c)\right]$ where $n_c=(N+1)/2$ is the lattice center and $\gamma>0$. Since $V_{\beta}=(-1)^{2n_c}V^*_{1-\beta}=(-1)^{2n_c} V_{1+\beta}$, it is sufficient to consider the family of potentials with $0<\beta< 1$ (when $\beta=0$ the problem reduces to the Aubrey-Andre model~\cite{harper,aa}). We then consider the effect of two such potentials $V_{\beta_1}+V_{\beta_2}$ with $|\beta_1-\beta_2|\sim 1/N\ll 1$.  

Our salient results are follows: i) For a single potential $V_\beta$, the threshold loss-gain strength $\gamma_{PT}(N,V_0,\beta)$ shows $N$ local maxima along the $\beta$ axis; it is suppressed by a nonzero real modulation $V_0$. ii) The discrete index of pair of eigenvalues that become complex can be tuned stepwise by varying $0<\beta< 1/2$. iii) For $V=V_{\beta_1}+V_{\beta_2}$, generically, the phase diagram in the $(\gamma_1,\gamma_2)$ plane shows a re-entrant $\mathcal{PT}$-symmetric phase: a broken $\mathcal{PT}$-symmetry is restored by increasing the non-hermiticity and broken again when $\gamma_i$ become sufficiently large. This behavior is absent in the extensively studied continuum Hamiltonians with complex potentials~\cite{makrispt,longpt,graefept}, and is a result of competition between the two lattice potentials $V_{\beta_1}$ and $V_{\beta_2}$.  

We emphasize that $\gamma_i>0$ means the gain-regions of the two potentials mostly align as do their respective loss-regions. Thus, the competition between $V_{\beta_1}$ and $V_{\beta_2\approx \beta_1}$ is not in their loss-gain profiles, but, as we will show below, due to the relative locations of $\mathcal{PT}$-symmetry breaking energy-levels in the spectrum. 

\begin{figure*}[thpb]
\begin{center}
\includegraphics[angle=0,width=0.9\columnwidth]{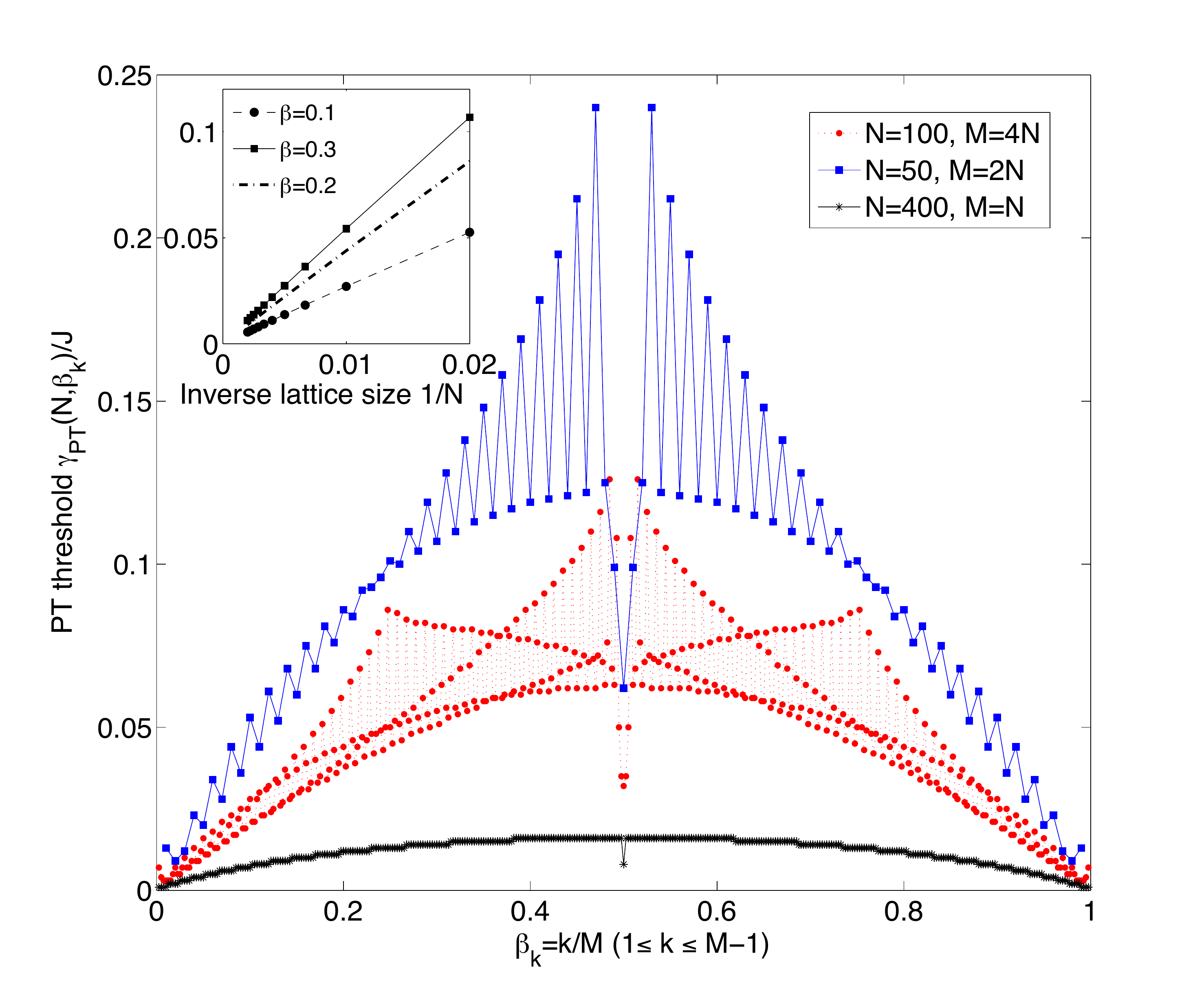}
\includegraphics[angle=0,width=0.94\columnwidth]{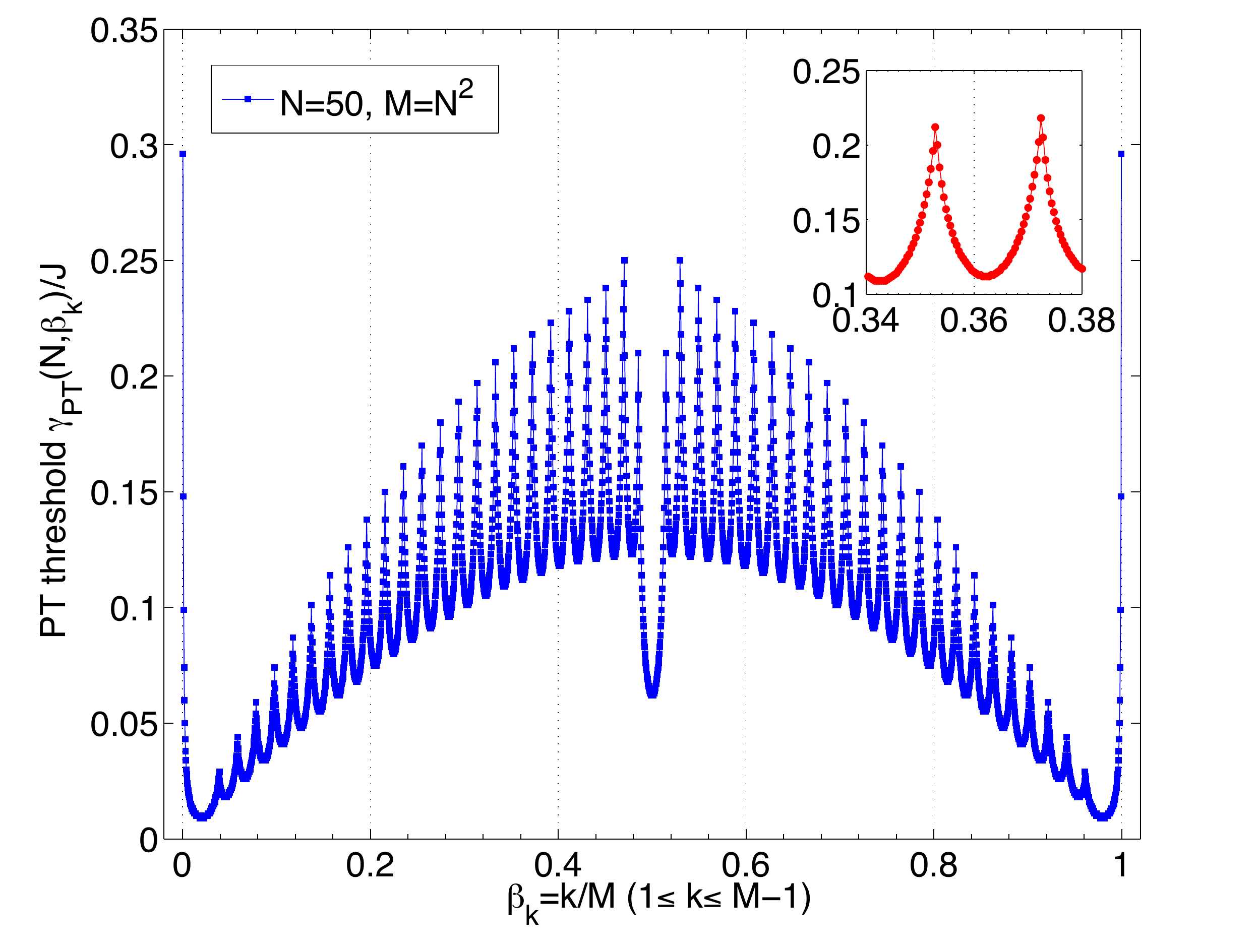}
\caption{(color online) Left-hand panel: $\mathcal{PT}$-symmetric threshold $\gamma_{PT}(N,\beta)$ for $N=50$ lattice with discretization $\beta_k=k\delta\beta=k/2N$ (blue squares), $N=100$ lattice with $\delta\beta=1/4N$ (red markers), and $N=400$ lattice with $\delta\beta=1/N$ (black stars) shows that the phase-diagram depends sensitively on $\delta\beta\sim O(1/N)$. The inset shows that for a fixed $\beta$, the threshold is linearly suppressed, $\gamma_{PT}(N,\beta)=C_\beta/N$. Right-hand panel: Results are independent of the discretization when $\delta\beta\sim 1/N^2\ll1/N$ and the $N=50$ phase-diagram shows $N$ local maxima and smooth oscillations with period $\Delta\beta=1/N$ (inset).}
\label{fig:phasediagram}
\end{center}
\vspace{-5mm}
\end{figure*}

\noindent{\it $\mathcal{PT}$ phase diagram for a single potential.} The tight-binding hopping Hamiltonian for an $N$-site lattice with open boundary conditions is $H_0=-J\sum_{n=1}^{N-1} (|n\rangle\langle n+1| + |n+1\rangle\langle n|)$. Its particle-hole symmetric energy spectrum is given by $\epsilon_{0,p}=-2J\cos(k_p)=-\epsilon_{0,\bar{p}}$ and the corresponding normalized eigenfunctions are $\psi_p(j)=\sin(k_pj)=(-1)^j\sin(k_{\bar{p}}j)$ where $0<k_p=p\pi/(N+1)<\pi$ and $\bar{p}=N+1-p$. The properties that relate eigenvalues and eigenfunctions at indices $p,\bar{p}$ remain valid in the presence of pure loss-gain potential $V_\beta=-V^*_\beta$~\cite{ph}. The eigenvalue equation for an eigenfunction $f(n)$ of the non-Hermitian, $\mathcal{PT}$-symmetric Hamiltonian $H_\beta=H_0+V_\beta$ is given by ($1\leq n\leq N$) 
\begin{equation}
\label{eq:tb}
-J\left[f(n+1)+f(n-1)\right]+ V_\beta(n)f(n)=E f(n), 
\end{equation} 
with $f(0)=0=f(N+1)$. Since this difference equation is not analytically soluble for an arbitrary $\beta$, we numerically obtain the spectrum $E(\gamma)$ and the $\mathcal{PT}$-symmetry breaking threshold $\gamma_{PT}(N,V_0,\beta)$ using different discretizations $\beta_k=k\delta\beta$ along the $\beta$-axis. Due to the $\beta\leftrightarrow 1-\beta$ symmetry of the potential, it follows that the exact threshold loss-gain strength satisfies $\gamma_{PT}(N,V_0,\beta)=\gamma_{PT}(N,V_0,1-\beta)=\gamma_{PT}(N,-V_0,\beta)$.
 
We consider a purely loss-gain potential, present results for an even lattice, and point out the salient differences that arise when lattice size $N$ is odd or when $V_0\neq 0$. The left-hand panel in Fig.~\ref{fig:phasediagram} shows the $\mathcal{PT}$-symmetric threshold $\gamma_{PT}(N,\beta)/J$ for an $N=50$ lattice obtained by using discretization $\delta\beta=1/2N$ (blue squares), an $N=100$ lattice with $\delta\beta=1/4N$ (red markers), and an $N=400$ lattice with $\delta\beta=1/N$ (black stars). There is a monotonic suppression of the threshold strength with increasing $N$, and, crucially, the general shape of the phase diagram depends upon the size of $\delta\beta$ relative to $1/N$. A scaling of this threshold suppression for lattice sizes $N=50-500$, shown in the inset, implies that $\gamma_{PT}(N,\beta)=C_\beta/N$ where $C_\beta$ is a constant. Thus, the threshold strength is suppressed linearly and vanishes in the thermodynamic limit~\cite{mark,dima}. However, this algebraically fragile nature of the $\mathcal{PT}$-phase is not an impediment since $\mathcal{PT}$-systems to-date are only realized in small lattices with $N\ll 100$. The right-hand panel in Fig.~\ref{fig:phasediagram} shows the phase diagram for $N=50$ case with discretization $\delta\beta=1/N^2$. The results for irrational values of $\beta$ and other discretizations lie on the same curve. The $\mathcal{PT}$ phase diagram shows $(N-2)$ local maxima located at $\beta_k=(2k+1)/2N$ and the two maxima at the end points, is symmetric about the center and has a local minimum in the threshold at $\beta=1/2$. In addition, the function $\gamma_{PT}(N,\beta)$ has $(N-1)$ minima at $\beta_k=k/N$ and smoothly oscillates over a period $\sim1/N=0.02$ as shown in the inset (solid red circles). These results are generic for any lattice size $N$ with discretization $\delta\beta\sim 1/N^2$. 

When $N$ is odd, the non-Hermitian potential vanishes at $\beta=1/2$ and the spectrum of the Hamiltonian $H_\beta$ is purely real. For an odd lattice, a similarly obtained phase diagram shows $(N-1)$ local maxima that are distributed equally on the two sides of $\beta=1/2$, along with a substantial enhancement in the threshold strength as $\beta\rightarrow 1/2^{\pm}$. Adding a real potential modulation $V_0\neq 0$, to the loss-gain potential, in general, suppresses the threshold strength. 

The phase diagram can be understood as follows: for a small $\beta\sim 1/N^2\ll 1/N$, $V_\beta(n)=i\gamma (2\pi\beta) (n-n_c)$ and the enhanced $\mathcal{PT}$-breaking threshold, $\gamma_{PT}/J\sim 0.3$, is consistent with a linear-potential threshold~\cite{microbridge}. For an even lattice, the average of the gain-potential is given by $A_E(\beta)=\sum_{n>n_c}^N V_\beta(n)/i\gamma=\sin^2(N\pi\beta/2)/\sin(\pi\beta)\geq 0$. The $\mathcal{PT}$ threshold is greatest when the {\it change in the average strength is maximum} as $\beta$ is varied, $\partial_\beta^2 A_E(\beta)=0$. In the limit $N\gg 1$ and $\beta\gg1/N^2$, it implies that the $N$ maxima of $\gamma_{PT}(N,\beta)$ occur at $\beta_{k,\max}=(2k+1)/2N$. On the other hand, $\gamma_{PT}(N,\beta)$ is smallest when the {\it change in the average strength is minimum}, $\partial_\beta A_E(\beta)=0$, and gives the locations of $(N-1)$ minima as $\beta_{k,\min}=k/N$. A similar analysis applies to odd lattices, where the average potential is given by $A_O(\beta)=\sin[\pi\beta(N-1)/2]\sin[\pi\beta(N+1)2]/\sin(\pi\beta)$. 

\begin{figure}[bthp]
\begin{center}
\includegraphics[angle=0,width=0.95\columnwidth]{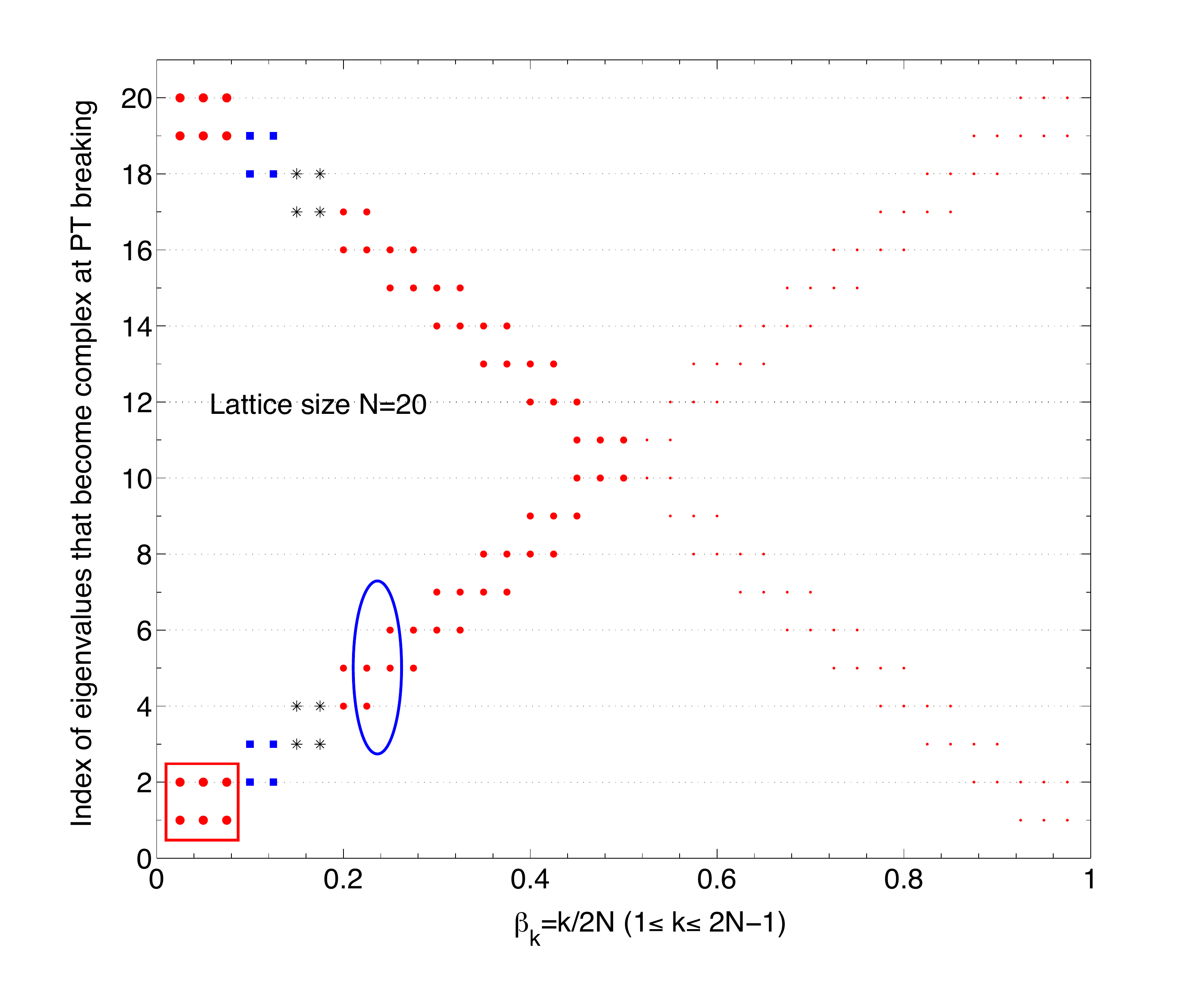}
\caption{(color online)  Index of eigenvalues that become complex as a function of $\beta$ for an $N=20$ lattice with discretization $\delta\beta=1/2N=0.025$ shows a $\beta\leftrightarrow 1-\beta$ symmetry, denoted by heavy and light red markers. When $\beta\leq 0.08$, levels $(E_1,E_2)$ become degenerate and complex, and so do their particle-hole counterparts, $(E_{19},E_{20})$ (red circles); in general, we can tune the location of $\mathcal{PT}$-breaking (blue squares, black stars, red markers) by appropriately choosing $\beta$.} 
\label{fig:ptlocation}
\end{center}
\vspace{-5mm}
\end{figure}
Next, we focus on the {\it location} of the $\mathcal{PT}$ symmetry breaking. Due to the particle-hole symmetric spectrum of $H_\beta$, two pairs of levels $(E_n,E_{n+1})$ and $(-E_n, -E_{n+1})$ become complex simultaneously. Figure~\ref{fig:ptlocation} plots the indices of eigenvalues that become complex as a function of $\beta$ for an $N=20$ lattice with discretization $\delta\beta=1/2N$. It shows that at small $\beta$, the eigenvalues at the band edges become complex, whereas, as $\beta\rightarrow 1/2$, the pairs of eigenvalues that become complex move to the center of the band. Thus the average range of $\beta$s with the same location for $\mathcal{PT}$-symmetry breaking is $\sim 1/N$. It follows that by choosing an appropriate $\beta$, one is able to control the location of $\mathcal{PT}$-symmetry breaking in the energy spectrum.  As we will see next, this control allows us to introduce competition between potentials $V_\beta$ with two different, but close, values of $\beta$.


\begin{figure}[bthp]
\begin{center}
\includegraphics[angle=0,width=0.9\columnwidth]{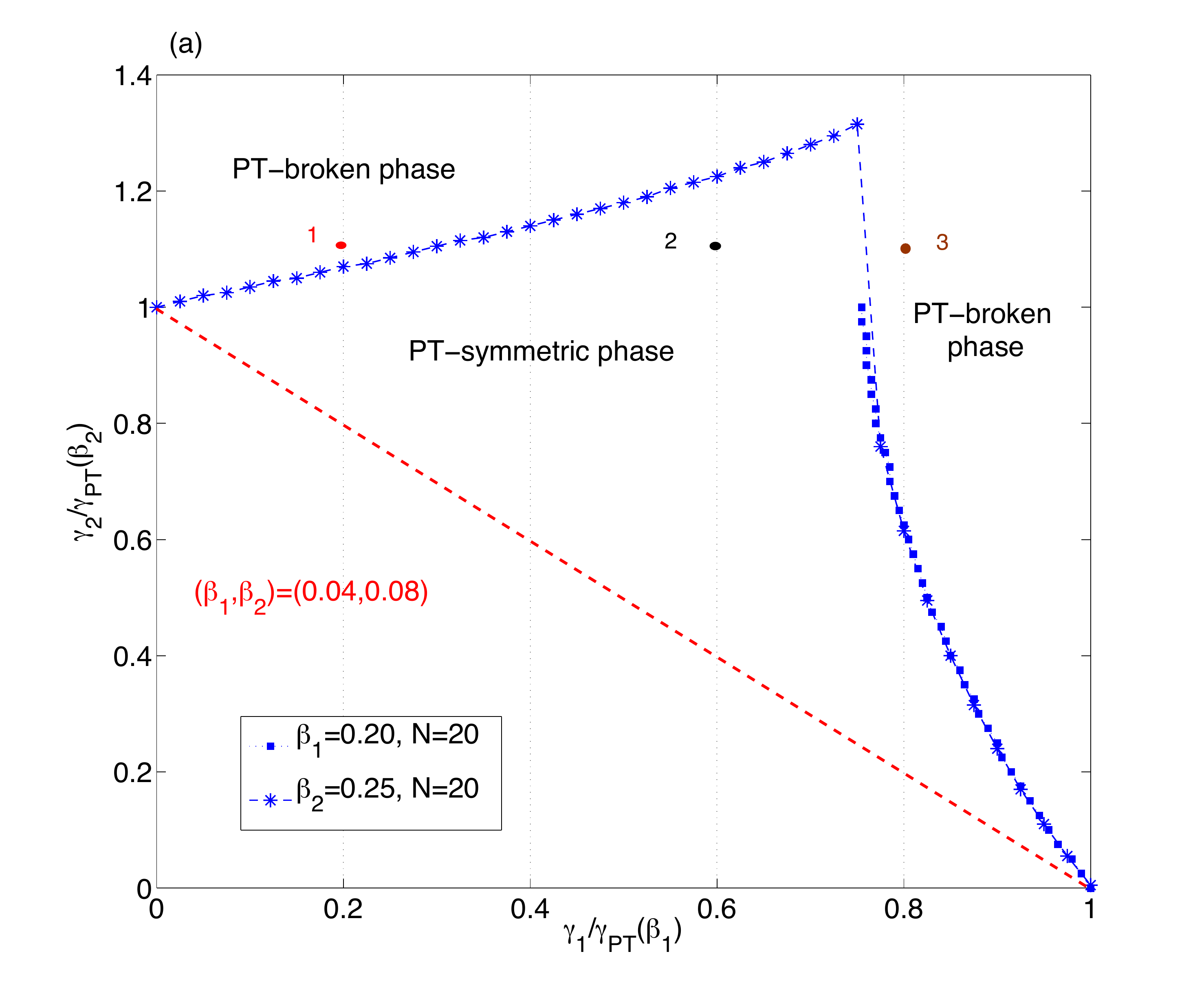}
\includegraphics[angle=0,width=0.95\columnwidth]{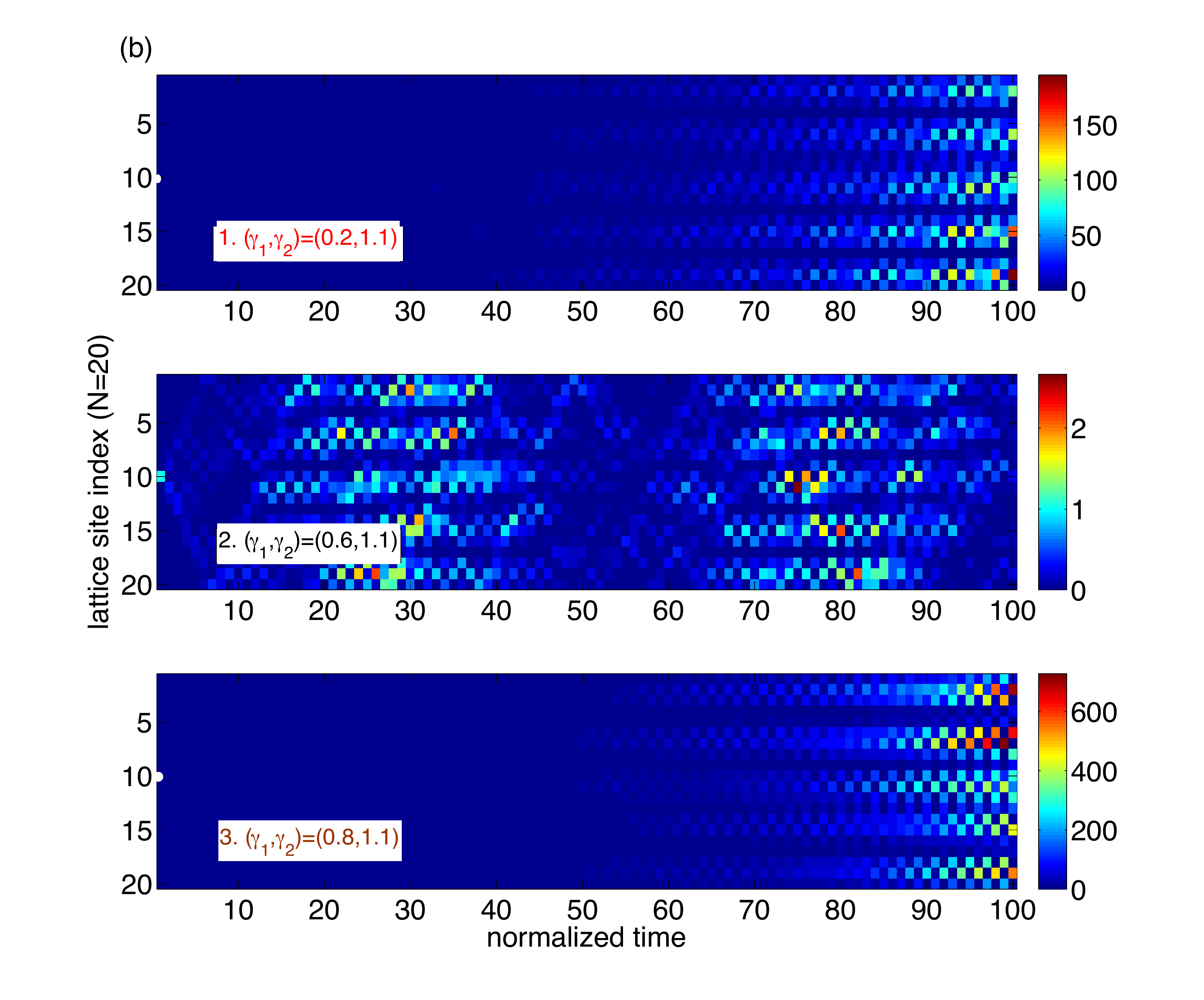}
\caption{(color online) Panel (a): $\mathcal{PT}$-phase diagram for potential $V_{\beta_1}+V_{\beta_2}$. When $(\beta_1,\beta_2)=(0.20,0.24)$ (blue stars and squares), a $\mathcal{PT}$-broken phase (point 1) is restored by increasing $\gamma_1$ (point 2) and subsequently broken again (point 3). This re-entrant phase is due to competition between $V_{\beta_1}$ and $V_{\beta_2}$. For $(\beta_1,\beta_2)=(0.04,0.08)$, the two co-operate and the phase boundary is an expected straight line. Panel (b): intensity $I(k,t)$ of an initially normalized state shows that, starting from a $\mathcal{PT}$-broken phase (top panel), increasing $\gamma_1$ initially restores bounded oscillations (center panel), followed by $\mathcal{PT}$ breaking and amplification (bottom panel). Note the two-orders-of-magnitude difference in the total intensity.}
\label{fig:pt1pt2}
\end{center}
\vspace{-5mm}
\end{figure}
\noindent{\it $\mathcal{PT}$ phase diagram with two potentials.} We now consider the $\mathcal{PT}$-symmetric phase of the Hamiltonian with two potentials, $H=H_0+V_{\beta_1}+V_{\beta_2}$, in the $(\gamma_1,\gamma_2)$ plane, where both axes are scaled by their respective threshold values $\gamma_{jPT}=\gamma_{PT}(\beta_j)$. Panel (a) in Fig.~\ref{fig:pt1pt2} shows the numerically obtained phase diagram for an $N=20$ lattice with $(\beta_1,\beta_2)=(0.20,0.25)$ (blue stars and squares). It shows that, from a $\mathcal{PT}$-broken phase (point 1), it is possible to enter the $\mathcal{PT}$-symmetric phase by increasing the non-hermiticity $\gamma_1$ (point 2). We emphasize that increasing $\gamma_1$ increases the average gain- (and loss-) strength $\gamma_1A_E(\beta_1)+\gamma_2 A_E(\beta_2)$, and yet drives the system into a $\mathcal{PT}$-symmetric phase from a $\mathcal{PT}$-broken phase. Increasing $\gamma_1$ further, eventually, drives the system into a $\mathcal{PT}$ broken phase again (point 3). 

While a $\mathcal{PT}$-broken phase implies an exponential time-dependence of the net intensity or the norm of a state, in the $\mathcal{PT}$-symmetric phase the net intensity oscillates within a bound that is determined by the proximity of the Hamiltonian to the $\mathcal{PT}$ phase boundary~\cite{review2}. Panel (b) in Fig.~\ref{fig:pt1pt2} shows the dramatic consequences of $\mathcal{PT}$ restoration on the site- and time-dependent intensity $I(k,t)=|\langle k|\exp(-iHt/\hbar)|\psi_0\rangle|^2$ of a state $|\psi_0\rangle$ that is initially localized on site $N/2=10$; the time is measured in units of $\hbar/J$. The top-subpanel shows that intensity has a monotonic amplification in regions with gain sites, leading to a striated pattern (point 1). Center subpanel shows that by increasing $\gamma_1$, oscillatory  behavior in the intensity is restored (point 2).  Bottom subpanel shows that increasing $\gamma_1$ further breaks the  $\mathcal{PT}$ symmetry again (point 3). Thus, we are able to {\it restore $\mathcal{PT}$-symmetry by increasing the non-hermiticity}, and achieve amplification by both reducing or increasing the average gain-strength. This novel behavior is absent in all lattice models with a single $\mathcal{PT}$ potential. 

Since each potential $V_{\beta_j}$ breaks the $\mathcal{PT}$ symmetry for $\gamma_j/\gamma_{jPT}>1$, naively, one may expect that the $\mathcal{PT}$ phase boundary for $V_{\beta_1}+V_{\beta_2}$ is given by $\gamma_1/\gamma_{1PT}+\gamma_2/\gamma_{2PT}=1$. This is indeed the case for $(\beta_1,\beta_2)=(0.04,0.08)$, shown by red dashed line in 
panel (a), even though the potential periodicities differ by a factor of two. 

What is the key difference between the two sets of parameters, one of which shows a re-entrant $\mathcal{PT}$-symmetric phase? It is the indices of eigenvalues that become complex due to $V_{\beta_1}$ and $V_{\beta_2}$. The red rectangle in Fig.~\ref{fig:ptlocation} shows that for $\beta\leq 0.8$, eigenvalues $(E_1,E_2)$ become complex. In such a case, the two potentials $V_{\beta_1}$ and $V_{\beta_2}$ act in a cooperative manner effectively adding their strengths. Therefore, the $\mathcal{PT}$-phase boundary is a straight line. In contrast, the blue oval in Fig.~\ref{fig:ptlocation} shows that for $\beta_1=0.20$, energy levels $E_4, E_5$ approach each other, become degenerate, and then complex as $\gamma\rightarrow\gamma_{1PT}$; for $\beta_2=0.25$, the energy levels that become complex as $\gamma\rightarrow\gamma_{2PT}$ are $E_5, E_6$. Thus, the level $E_5$ is {\it lowered by potential $V_{\beta_1}$ and raised by the potential $V_{\beta_2}$} from its hermitian-limit value. This introduces competition between the two potentials $V_{\beta_1}$ and $V_{\beta_2}$ even though their gain-regions largely overlap and so do their respective loss regions.  

\begin{figure}[bthp]
\begin{center}
\includegraphics[angle=0,width=1\columnwidth]{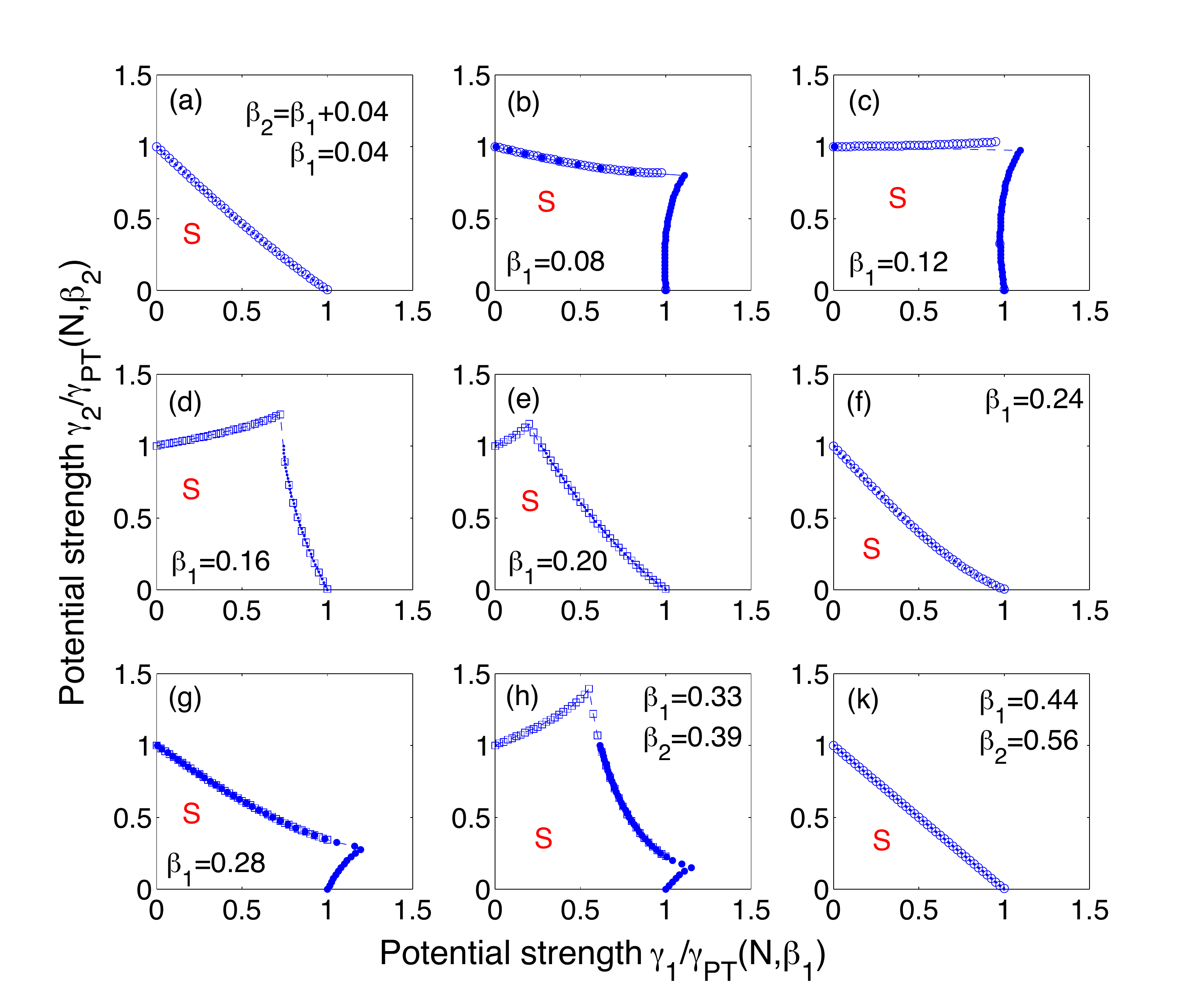}
\caption{(color online) $\mathcal{PT}$-phase boundary for an $N=20$ lattice with potentials $V_{\beta_1}+V_{\beta_2}$ shows that $\mathcal{PT}$-symmetry restoration can occur by increasing non-hermiticity $\gamma_1$ (panels b, c, g), $\gamma_2$ (panels d, e), or both (panel h) when the two potentials compete. When they make complex the same set of eigenvalues, the phase boundary is a line (panels a, f, k). The label ``S'' denotes the $\mathcal{PT}$-symmetric phase.}
\label{fig:ptglobal}
\end{center}
\vspace{-5mm}
\end{figure}

This correspondence between completing potentials and $\mathcal{PT}$-restoration is further elucidated in Fig.~\ref{fig:ptglobal}. In conjunction with Fig.~\ref{fig:ptlocation}, it shows that re-entrant $\mathcal{PT}$-symmetric phase occurs when the two potentials compete (panels b-e, g, h). This restoration of $\mathcal{PT}$-symmetry can be due to increased loss-gain strength in $\gamma_1$ (panels d, e), $\gamma_2$ (panels b, c, g), or both (panel h). On the other hand, when the two potentials break the same set of eigenvalues, the $\mathcal{PT}$ phase boundary is a line (panels a, f, k). 

\noindent{\it Discussion.}  Competing potentials, a common theme in physics, often stabilize phases that would be unstable in the presence of only one of them~\cite{fradkin}. A trivial definition of competing $\mathcal{PT}$-potentials is that the gain-region of one strongly overlaps with the loss-region of another, thus reducing the average gain (and loss) strength. 

Here, we have unmasked the subtle competition between $\mathcal{PT}$ potentials whose gain regions largely overlap, based on the location of $\mathcal{PT}$-symmetry breaking induced by each. This competition results in $\mathcal{PT}$-restoration and subsequent $\mathcal{PT}$-breaking, leading to selective intensity suppression and oscillations at large loss-gain strength. Its hints were seen in a continuum model with complex $\delta$-function and constant potentials, but that continuum model is neither easily experimentally realizable nor can it tune between cooperative and competitive behavior~\cite{bijan}. The $\mathcal{PT}$-symmetric Aubry-Andre model provides a family of potentials with tunable competition or cooperation among them, and is thus ideal for investigating the consequences of such competition; even lattices as small as $N=10$ that can be realized via coupled optical waveguides~\cite{review1,review2} or cold atoms~\cite{huang} may provide a comprehensive understanding of interplay between loss-gain strengths and $\mathcal{PT}$-symmetry breaking. 


This work was supported by NSF Grant No. DMR-1054020. 


\end{document}